\documentclass[entropy,article,accept,moreauthors,pdftex,10pt,a4paper]{mdpi} 

\usepackage{graphicx}
\usepackage{amssymb}
\usepackage{epstopdf}
\usepackage{bbm}
\usepackage{mathtools}
\usepackage{physics}

\firstpage{1} 
\articlenumber{x}
\doinum{10.3390/------}
\pubvolume{xx}
\pubyear{2017}
\copyrightyear{2017}
\history{Received: 9 October 2017; Accepted: 20 November 2017; Published: date}

\pdfoutput=1
\usepackage{soul,upgreek}
\usepackage{booktabs} 
\usepackage{multirow}
\usepackage{soul} 
\usepackage{microtype}

\setitemize{parsep=6pt,itemsep=0pt,leftmargin=*,labelsep=5.5mm}
\setenumerate{parsep=6pt,itemsep=0pt,leftmargin=*,labelsep=5.5mm}
\setlist[description]{itemsep=0mm}

 \theoremstyle{mdpi}
 \newcounter{thm}
 \newcounter{ex}
 \newcounter{re}
\newcommand{\spin}{\boldsymbol{s}}

\newcommand{\vecrho}{\boldsymbol{\varrho}}

\newcommand{\vecv}{\boldsymbol{v}}
\newcommand{\vecu}{\boldsymbol{u}}

\newcommand{\vecg}{\boldsymbol{g}}

\newcommand{\partf}{\mathcal{Z}}

\Title{On Maximum Entropy and Inference}

\Author{Luigi Gresele $^{1,2}$ and Matteo Marsili $^{3,}$*}

\AuthorNames{Firstname Lastname, Firstname Lastname and Firstname Lastname}

\address{%
$^{1}$ \quad Max Planck Institute for Intelligent Systems, Empirical Inference Department, Max-Planck-Ring 4, 72076 T\"ubingen, Germany; luigi.gresele@tuebingen.mpg.de\\
$^{2}$ \quad Max Planck Institute for Biological Cybernetics, 
High-field Magnetic Resonance Department, Max-Planck-Ring 11, 72076 T\"ubingen, Germany\\
$^{3}$ \quad The Abdus Salam International Center for Theoretical Physics, Quantitative Life Sciences Section, Strada Costiera 11, 34151 Trieste, Italy}

\corres{Correspondence: marsili@ictp.it; Tel.: +39-040-224-0461}

\abstract{maximum entropy is a powerful concept that entails a sharp separation between relevant and irrelevant variables. 
It is typically invoked in inference, once an assumption is made on what the relevant variables are, in order to estimate a model from data, that affords predictions on all other (dependent) variables. 
Conversely, maximum entropy can be invoked to retrieve the relevant variables (sufficient statistics) directly from the data, once a model is identified by Bayesian model selection. We explore this approach in the case of spin models with interactions of arbitrary order, and we discuss how relevant interactions can be inferred. In this perspective, the dimensionality of the inference problem is not set by the number of parameters in the model, but by the frequency distribution of the data. We illustrate the method showing its ability to recover the correct model in a few prototype cases and discuss its application on a real dataset.}


\begin{document}




\section{Introduction}

Statistical mechanics stems from classical (or quantum) mechanics. The latter prescribes which are the relevant quantities (i.e., the conserved ones). The former brings this further, and it predicts that the probability to observe a system in a microscopic state $\spin$, in thermal equilibrium, is given by: 
	\begin{equation}
	\label{Boltzmann}
	P(\spin)=\frac{1}{\partf}e^{-\beta H[\spin]}
	\end{equation}
	where $H[\spin]$ is the energy of configuration $\spin$. The inverse temperature $\beta$ is the only relevant parameter that needs to be adjusted, so that the ensemble average $\langle H\rangle$ matches the observed energy $U$. It has been argued \cite{jaynes1957information} that the recipe that leads from $H[\spin]$ to the distribution $P(\spin)$ is maximum entropy: {among all distributions that satisfy $\langle H\rangle=U$, the one maximizing the entropy $S=-\sum_{\spin} P(\spin)\log P(\spin)$ should be chosen}.
	Information theory clarifies that the distribution Equation (\ref{Boltzmann}) is the one that assumes nothing else but $\langle H\rangle=U$, or equivalently, that all other observables can be {predicted} from the knowledge of~$H[\spin]$.
	
	This idea carries through more generally to inference problems: given a dataset of $N$ observations $\hat s=\{\spin^{(1)},\ldots,\spin^{(N)}\}$ of a system, one may invoke maximum entropy to infer the underlying distribution $P(\spin)$ that reproduces the empirical averages of a set $\mathcal{M}$ of observables $\phi^\mu(\spin)$ ($\mu\in \mathcal{M}$). 
	This leads to Equation (\ref{Boltzmann}) with:
	\begin{equation}
	\label{EqH}
	-\beta H[\spin]=\sum_{\mu\in \mathcal{M}}g^\mu\phi^\mu(\spin)
	\end{equation}
	where the parameters $g^\mu$ should be fixed by solving the convex optimization problems: 
	\begin{equation}
	\label{eq:inference}
	\langle \phi^\mu\rangle=\overline{\phi^\mu}\equiv \frac{1}{N}\sum_{i=1}^N\phi^\mu\left(\spin^{(i)}\right)
	\end{equation}
	that result from entropy maximization and are also known to coincide with maximum likelihood estimation (see \cite{Pitman,Darmois,Koopman}). 
  
  For example, in the case of spin variables $\spin\in\{\pm 1\}^n$, the distribution that reproduces empirical averages $\overline{s_i}$ and correlations $\overline{s_is_j}$ is the pairwise model: 
	\begin{equation}
	\label{eq:pwise}
	H[\spin]=-\sum_i h_i s_i-\sum_{i<j} J_{ij} s_is_j\,,
	\end{equation}
  which in the case $J_{ij}\equiv J$ $\forall i,j$ and $h_i\equiv h$ $\forall i$ is the celebrated Ising model. The literature on inference of Ising models, stemming from the original paper on {Boltzmann learning} \cite{Hinton} to early applications to neural data \cite{Bialek1} has grown considerably (see \cite{review} for a recent review), to the point that some suggested~\cite{BialekUSSC} that a purely data-based statistical mechanics is possible.
	
	Research has mostly focused on the estimate of the parameters $\vecg=\{h_i,J_{ij}\}$, which itself is a computationally challenging issue when $n\gg 1$ \cite{review}, or in recovering sparse models, distinguishing true interactions ($J_{ij}\neq 0$) from spurious ones ($J_{ij}=0$; see, e.g., \cite{wainwright2003variational}). Little has been done to go beyond pairwise interactions (yet, see \cite{Higher-order-Sejnowski,Amari3,Margolin2010,Ilya}).
	This is partly because pairwise interactions offer a convenient graphical representation of statistical dependences; partly because $\ell$-th order interactions require $\sim$n$^\ell$ parameters and the available data hardly ever allow one to go beyond $\ell=2$ \cite{footnote1}. 
	 	Yet, strictly speaking, there may be no reason to believe that interactions among variables are only pairwise. The choice of form (\ref{eq:pwise}) for the Hamiltonian represents an assumption on the intrinsic laws of motion, which reflects an {a priori} belief of the observer on the system. Conversely, one would like to have inference schemes that certify that pairwise interactions are really the {relevant} ones, i.e., those that need to be included in $H$ in order to reproduce correlations $\overline{s_{i_1}s_{i_2}\cdots s_{i_\ell}}$ of arbitrary order $\ell$ \cite{footnote2}.
	
	We contrast a view of inference as parameter estimation of a preassigned (pairwise) model, where~maximum entropy serves merely an ancillary purpose, with the one where the ultimate goal of statistical inference is precisely to identify the minimal set $\mathcal{M}$ of sufficient statistics that, for a given dataset $\hat s$, accurately reproduces all empirical averages. In this latter perspective, maximum entropy plays a key role in that it affords a sharp distinction between {relevant} variables ($\phi^\mu(\spin)$, $\mu\in \mathcal{M}$) (which~are the sufficient statistics) and {irrelevant} ones, i.e., all other operators that are not a linear combination of the relevant ones, but whose values can be predicted through theirs. To some extent, understanding amounts precisely to distinguishing the relevant variables from the ``dependent'' ones: those whose values can be predicted. 
	
	Bayesian model selection provides a general recipe for identifying the best model $\mathcal{M}$; yet, as we shall see, the procedure is computationally unfeasible for spin models with interactions of arbitrary order, even for moderate dimensions ($n=5$). Our strategy will then be to perform model selection within the class of mixture models, where it is straightforward \cite{marsilihaimovici}, and then, to project the result on spin models. The most likely models in this setting are those that enforce a symmetry among configurations that occur with the same frequency in the dataset. These symmetries entail a decomposition of the log-likelihood with a flavor that is similar to principal component analysis (yet of a different kind than that described in \cite{principal-value-dec}). This 
	directly predicts the sufficient statistics $\psi_\lambda(\spin)$ that need to be considered in maximum entropy inference. Interestingly, we find that the number of sufficient statistics depends on the frequency distribution of observations in the data. This implies that the dimensionality of the inference problem is not determined by the number of parameters in the model, but rather by the richness of the data. 
  
  The resulting model features interactions of arbitrary order, in general, but is able to recover sparse models in simple cases. An application to real data shows that the proposed approach is able to spot the prevalence of two-body interactions, while suggesting that some specific higher order terms may also be important.

\section{Background}
\subsection{Spin Models with Interactions of Arbitrary Order}
	\label{arbitrary}
	
	Consider a system of $n$ spin variables $s_i=\pm 1$, a state of which is defined by a configuration $\spin=(s_1, \ldots, s_n)$. The number of all possible states is $2^n$. A generic model is written as:
	\begin{equation}
	\label{eq:1}
	P(\spin|\vecg,\mathcal{M})=\frac{1}{\partf}e^{\sum_{\mu\in\mathcal{M}} g^{\mu}\phi^{\mu}(\spin)}\,,
	\end{equation}
	where $\phi^{\mu}(\spin)=\prod_{i \in \mu}s_i$ is the product of all spins involved in the corresponding interaction $g^\mu$, the sum on $\mu$ runs on a subset of operators $\phi^{\mu}$ and $\partf$ ensures normalization. We follow the same notation as in \cite{Beretta}: There are $2^n$ possible such operators, which can be indexed by an integer $\mu=0,\ldots,2^n-1$ whose binary representation indicates those spins that occur in the operator $\phi^\mu$. Therefore, $\mu=0$ corresponds to the constant operator $\phi^0(\spin)=1$, and for $\mu=11$, the notation $i\in\mu$ is equivalent to $i\in\{1,2,4\}$, i.e., $\phi^{11}(\spin)=s_1 s_2 s_4$. 
	
	Given a dataset $\hat s$ of $N$ observations of $\spin$, assuming them to be i.i.d. draws from $P(\spin|\vecg,\mathcal{M})$, the parameters $g^\mu$ are determined by solving Equation (\ref{eq:inference}). Bayesian inference maintains that different models should be compared on the basis of the posterior $P\{\mathcal{M}|\hat s\}$ that can be computed integrating the likelihood over the parameters $\vecg$ (we refer to \cite{Beretta} for a discussion of Bayesian model selection within this setup). This can already be a daunting task if $n$ is large. Note that each operator $\phi^\mu$ ($\mu>0$) can either be present or not in $\mathcal{M}$; this implies that the number of possible models is $2^{2^n-1}$. Finding~the most likely model is impossible, in practice, even for moderately large $n$. 
	
	
	\subsection{Bayesian Model Selection on Mixture Models}
	\label{mixture}


Let us consider {mixture} models in which the probability of state $\spin$ is of the form:
	\begin{equation}
	\label{mixturecomb}
	P(\spin|\vecrho,\mathcal{Q})=\sum_{j=1}^q \rho_j \mathbbm{1}_{Q_j}(\spin)
	\end{equation}	
	where $\mathbbm{1}_A(\spin)$ is the indicator function $\mathbbm{1}_A(\spin)=1$ if $\spin \in A$, and $\mathbbm{1}_A(\spin)=0$ otherwise. The prior (and posterior) distributions of $\vecrho$ take a Dirichlet form; see Appendix \ref{Dirichlet}. In other words, model $\mathcal{Q}$ assigns the same probability $\rho_j$ to all configurations in the same set $Q_j$. Formally, $\mathcal{Q}=\{Q_i\}$ is a partition of the set of configurations $\spin$, i.e., a collection of subsets such that $\{\pm 1\}^n=\bigcup_j Q_j$ and $Q_j\bigcap Q_{j'}=\emptyset$ $\forall j\neq j'$. The~model's parameters $\rho_j$ are subject to the normalization constraint $\sum_j|Q_j|\rho_j=1$, where $|Q|$ stands for the number of elements within $Q$. We denote by $q=|\mathcal{Q}|$ the number of subsets in $\mathcal{Q}$. The number of independent parameters in model $\mathcal{Q}$ is then $q-1$.
	
The number of possible models $\mathcal{Q}$ is the number of partitions of a set of $2^n$ elements, which is the Bell number $B_{2^n}$. This grows even faster than the number of spin models $\mathcal{M}$. Yet, Bayesian model selection can be easily carried out, as shown in \cite{marsilihaimovici}, assuming Dirichlet's prior. In brief, 
the~most likely partition $\mathcal{Q}^*$ depends on the assumed prior, but it is such that if two states $\spin$ and $\spin'$ are observed a similar number of times $k_{\spin}\simeq k_{\spin'}$, then the most likely model places them in the same set $Q_q$ and assigns them the same probability $\rho_q$. In other words, considering the {frequency partition}:
\begin{equation}
\label{ }
\mathcal{K}=\{Q_k\},\qquad Q_k=\{s:~k_s=k\}\,,
\end{equation}
that groups in the same subset all states $s$ that are observed the same number $k_s$ of times, the {optimal partition} $\mathcal{Q}^*$ is always a coarse graining of $\mathcal{K}$, likely to merge together subsets corresponding to similar empirical frequencies.

For example, in the $n=1$ case, the possible partitions are $\mathcal{Q}^{(1)}=\{\{\pm 1\}\}$ and $\mathcal{Q}^{(2)}=\{\{- 1\},\{+ 1\}\}$. The first corresponds to a model with no parameters where $P(s|\mathcal{Q}^{(1)})=1/2$ for $s=\pm 1$, whereas the second assigns probability $P(s=+1|\mathcal{Q}^{(2)})=\rho_+$ and $P(s=-1|\mathcal{Q}^{(2)})=1-\rho_+$. Detailed calculation \cite{marsilihaimovici} shows that model $\mathcal{Q}^{(1)}$ should be preferred unless 
$\overline{s}$ is sufficiently different from zero (i.e., the frequency with which the states $s=\pm 1$ are observed is sufficiently different). 

We refer the interested reader to Appendix \ref{Dirichlet} and \cite{marsilihaimovici} for more details, as well as for a heuristic for finding the $\mathcal{Q}^*$ model. 
	
\section{Mapping Mixture Models into Spin Models}
\label{mapping}
	
Model $\mathcal{Q}$ allows for a representation in terms of the variables $g^\mu$, thanks to the relation:
\begin{equation}
\label{eq:mapp}
g^\mu_{\mathcal{Q}}=\frac{1}{2^n}\sum_{\spin}\phi^\mu(\spin)\log P(\spin|\vecrho,\mathcal{Q})=
\frac{1}{2^n}\sum_{j}\chi^\mu_j\log \rho_j,\qquad \chi^\mu_j=\sum_{\spin\in Q_j}\phi^\mu(\spin)\,,
\end{equation}
which is of the same nature of the one discussed in \cite{Amari3} 
and whose proof is deferred to Appendix~\ref{appendix:comp}. 
The index in $g^\mu_{\mathcal{Q}}$ indicates that the coupling refers to model $\mathcal{Q}$ and merely corresponds to a change of variables $\vecrho \to \vecg$; we shall drop it in what follows, if it causes no confusion. 

In Bayesian inference, $\vecrho$ should be considered as a random variable, whose posterior distribution for a given dataset $\hat s$ can be derived (see \cite{marsilihaimovici} and Appendix \ref{Dirichlet}). Then, Equation (\ref{eq:mapp}) implies that also $\vecg$ is a random variable, whose distribution can also be derived from that of $\vecrho$.

Notice, however, that Equation (\ref{eq:mapp}) spans only a $q-1$-dimensional manifold in the $2^n-1$-dimensional space $\vecg$, because there are only $q-1$ independent variables $\vecrho$. This fact is made more evident by the following argument: Let $\vecv$ be a $2^n-1$ component vector such that: 
\begin{equation}
\label{eq:v}
\sum_{\mu=1}^{2^n-1} v^\mu\chi^\mu_j=0,\qquad\forall j=1,\ldots,q.
\end{equation}

Then, we find that:
\begin{equation}
\label{eq:gv}
\sum_\mu v^\mu g^\mu=0.
\end{equation}

In other words, the linear combination of the random variables $g^\mu$ with coefficients $v^\mu$ that satisfy Equation (\ref{eq:v}) is not random at all. There are (generically) $2^n-1-q$ vectors $\vecv$ that satisfy Equation (\ref{eq:v}) each of which imposes a linear constraint of the form of Equation (\ref{eq:gv}) on the possible values of $\vecg$. 

In addition, there are $q$ orthogonal directions $\vecu_\lambda$ that can be derived from the singular value decomposition of $\chi^\mu_j$:
\begin{equation}
\label{eq:svd}
\chi_j^\mu=\sum_{\lambda=1}^{q}\Lambda_\lambda u_\lambda^\mu w_{\lambda,j},\qquad \sum_\mu u_\lambda^\mu u_{\lambda'}^\mu=\delta_{\lambda,\lambda'},\qquad \sum_j w_{\lambda,j} w_{\lambda',j}=\delta_{\lambda,\lambda'}.
\end{equation}

This in turn implies that model $\mathcal{Q}$ can be written in the exponential form (see Appendix \ref{eigsuff} for details):
\begin{equation}
\label{eq:maxentlamb}
P(\spin|\vecg,\mathcal{Q})=\frac{1}{\partf}e^{\sum_{\lambda=1}^{q}g_\lambda\psi_\lambda(\spin)},
\end{equation}
where:
\begin{align}
\label{eq:psi}
& \psi_\lambda(\spin)=\sum_\mu u_\lambda^\mu\phi^\mu(\spin)\,;\\
& g_\lambda = \sum_\mu u_\lambda^\mu g^\mu\,.
\end{align}

The exponential form of Equation (\ref{eq:maxentlamb}) identifies the variables $\psi_\lambda(\spin)$ with the sufficient statistics of the model. The maximum likelihood parameters $\hat g^{\lambda}$ can be determined using the knowledge of empirical averages of $\psi_\lambda(\spin)$ alone, solving the equations $\langle \psi_\lambda\rangle=\overline{ \psi_\lambda}$ for all $\lambda=1,\ldots,q$. The resulting distribution is the maximum entropy distribution that reproduces the empirical averages of $\psi_\lambda(\spin)$. In~this precise sense, $\psi_\lambda(\spin)$ are the relevant variables. Notice that, the variables $\psi_\lambda(\spin)$ are themselves an orthonormal set:
\begin{equation}
\label{ }
\sum_{\spin}\psi_\lambda(\spin)=0,\qquad \frac{1}{2^n}\sum_{\spin}\psi_\lambda(\spin)\psi_{\lambda'}(\spin)=\delta_{\lambda,\lambda'}.
\end{equation}

In particular, if we focus on the $\mathcal{K}$ partition of the set of states, the one assigning the same probability $\rho_k$ to all states $\spin$ that are observed $k$ times, we find that $P(\spin|\hat\vecg,\mathcal{Q})=k_{\spin}/N$ exactly reproduces the empirical distribution. 
This is a consequence of the fact that the variables $\hat g_\lambda$ that maximize the likelihood must correspond to the maximum likelihood estimates $\hat \rho_k=k/N$, via Equation (\ref{eq:mapp}). This~implies that 
the maximum entropy distribution Equation (\ref{eq:maxentlamb}) reproduces not only the empirical averages $\psi_\lambda(\spin)$, but also that of the operators $\phi^\mu(\spin)$ for all $\mu$. 
A direct application of Equation (\ref{eq:mapp}) shows that the maximum entropy parameters are given by the formula: 
\begin{equation}
\label{eq:directk}
\hat g^\mu_{\mathcal{K}}=\frac{1}{2^n}\sum_{\spin}\phi^\mu(\spin)\log\frac{k_{\spin}}{N}=\frac{1}{2^n}\sum_{k}\chi^\mu_k\log\frac{k}{N}.
\end{equation}

Similarly, the maximum likelihood parameters $\hat g_\lambda$ are given by:
\begin{equation}
\hat{g}_\lambda=\frac{1}{2^n}\sum_{\spin}
\psi_\lambda(\spin)\log{\frac{k_{\spin}}{N}}=
\frac{\Lambda_\lambda}{2^n}\sum_{k}
w_{\lambda,k}\log{\frac{k}{N}}.
\label{eq:hatglambda}
\end{equation}

Notice that, when the set $Q_0=\{\spin:~k_{\spin} =0\}$ of states that are not observed is not empty, all~couplings $\hat g^\mu$ with $\chi_0^\mu\neq 0$ diverge. Similarly, all $\hat g_\lambda$ with $w_{\lambda,0}\neq 0$ also diverge. We shall discuss later how to {regularize} these divergences that are expected to occur in the under-sampling regime (i.e.,~when~$N\le 2^n$).

It has to be noted that, of the $q$ parameters $\rho_q$, only $q-1$ are independent. Indeed, we find that one of the $q$ singular values $\Lambda_\lambda$ in Equation (\ref{eq:svd}) is practically zero. 
It is interesting to inspect the covariance matrix $C^{\mu,\nu}=E[\delta g^\mu\delta g^\nu]$ of the deviations $\delta g^\mu=g^\mu-E[g^\mu]$ from the expected values computed on the posterior distribution. We find (see Appendices \ref{covarmat} and \ref{eigsuff}) that $C^{\mu,\nu}$ has eigenvalues $\Lambda^2_\lambda$ along the eigenvectors $\vecu_\lambda$ and zero eigenvalues along the directions $\vecv$. The components $\lambda$ with the largest singular value $\Lambda_\lambda$ are those with the largest statistical error, so one would be tempted to consider them as ``sloppy'' directions, as in \cite{sloppy}. Yet, by Equation (\ref{eq:hatglambda}), the value of $\hat g_\lambda$ itself is proportional to $\Lambda_\lambda$, so~the relative fluctuations are independent of $\Lambda_\lambda$. Indeed ``sloppy'' modes appear in models that overfit the data, whereas in our case, model selection on mixtures ensures that the model $\mathcal{Q}^*$ does not overfit. This is why relative errors on the parameters $g_\lambda$ are of comparable magnitude. Actually,~variables~$\psi_\lambda$ that correspond to the largest eigenvalues $\Lambda_\lambda$ are the most relevant ones, since they identify the directions along which the maximum likelihood distribution Equation~(\ref{eq:maxentlamb}) tilts most away from the unconstrained maximal entropy distribution $P_0(\spin)=1/2^n$. A further hint in this direction is that Equation (\ref{eq:hatglambda}) implies that variables $\psi_\lambda(\spin)$ with the largest $\Lambda_\lambda$ are those whose variation across states $\spin$ typically correlates mostly with the variation of $\log k_{\spin}$ in the sample. 

Notice that the procedure outlined above produces a model that is sparse in the $\varrho$ variables, i.e., it depends only on $q-1$ parameters, where $q$ is, in the case of the $\mathcal{K}$ partition, the number of different values that $k_s$ takes in the sample.
Yet, it is not sparse in the $g^\mu$ variables. Many of the results that we have derived carry through with obvious modifications if the sums over $\mu$ are restricted to a subset $\mathcal{M}$ of putatively relevant interactions. Alternatively, the results discussed above can be the starting point for the approximate scheme to find sparse models in the spin representation. 

	
\section{Illustrative Examples}
	
	In the following, we present simple examples clarifying the effects of the procedure outlined~above. 
		
	\subsection{Recovering the Generating Hamiltonian from Symmetries: Two and Four Spins}
	\label{sec:twospin}
	
	As a simple example, consider a system of two spins. The most general Hamiltonian that should be considered in the inference procedure is:
	
	\begin{equation}
	\label{twoham}
	H^{\text{Inf}}[\spin]=g^1\cdot s_1+g^2\cdot s_2+g^3\cdot s_1 s_2.
	\end{equation}

	Imagine the data are generated from the Hamiltonian: 
	
	\begin{equation}
	\label{twogen}
	H^{\text{Gen}}[\spin]=J\cdot s_1 s_2\,.
	\end{equation}
	and let us assume that the number of samples is large enough, so that the optimal partition $\mathcal{Q}^*$~groups configurations of aligned spins $Q_==\{\spin :s_1=s_2\}$ distinguishing them from the configuration of unaligned ones $Q_{\not =}=\{\spin :s_1=-s_2\}$. 
	
	Following the strategy explained in Section \ref{arbitrary}, we observe that $\chi^\mu_==\chi^\mu_{\not =}=0$ for both $\mu=1$ and $2$. Therefore, Equation (\ref{eq:mapp}) implies $g^1=g^2=0$. Therefore, the $\mathcal{Q}^*$ model only allows for $g^3$ to be nonzero. 	
	In this simple case, symmetries induced by the $\mathcal{Q}^*$ model (i.e., $(s_1,s_2)\to (-s_1,-s_2)$) directly produce a sparse model where all interactions that are not consistent with them are set to zero.

\begin{figure}[H]
		\centering
		\begin{minipage}[b]{0.3\textwidth}
			\includegraphics[width=0.6\textwidth]{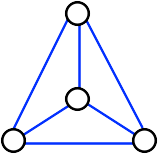}
			\caption{$\mathcal{G}_\text{Pw}$}
		\end{minipage}
		\hspace{5mm}
		\begin{minipage}[b]{0.3\textwidth}
			\includegraphics[width=0.6\textwidth]{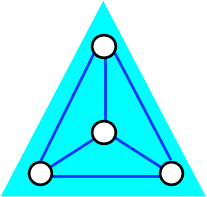}
			\caption{$\mathcal{G}_{\text{Pw} \cup \text{4-body}}$}
		\end{minipage}
		\caption{\label{fig:Spurious} A representation of the two models consistent with the $\mathcal{Q}^*$ partition in Eq. (\ref{eq:4spin_part}). Blue links represent pairwise interactions and the shaded area represents a four body interaction.}
	\end{figure}

	Consider now a four-spin system. Suppose that the generating Hamiltonian is that of a pairwise fully-connected model as in Figure \ref{fig:Spurious} (left), with the same couplings $g_3=g_5=g_6=g_9=g_10=g_12=J$. 
	With enough data, we can expect that the optimal model is based on the partition $\mathcal{Q^*}$ that distinguishes three sets of configurations:
\begin{equation}
	Q_j=\left\{\spin:~s_1+s_2+s_3+s_4=\pm 2j\right\},\qquad j=0,1,2
\label{eq:4spin_part}
\end{equation}	
depending on the absolute value of the total magnetization. The $\mathcal{Q^*}$ model assigns the same probability $\rho_j$ to configurations $\spin$ in the same set $Q_j$. 
	Along similar lines to those in the previous example, it can be shown that any interaction of order one is put to zero ($g^1=g^2=g^4=g^8=0$), as well as any interaction of order three ($g^7=g^{11}=g^{13}=g^{14}=0$), because the corresponding interactions are not invariant under the symmetry $\spin\to-\spin$ that leaves $\mathcal{Q}^*$ invariant. 
The interactions of order two will on the other hand correctly be nonzero and take on the same value $g_3=g_5=g_6=g_9=g_10=g_12$. The~value of the four-body interaction is: 
	\begin{equation}
	\label{four_body}
	g^{15}=\frac{1}{2^4}\left[ 2\cdot \log \rho_2 -2 \cdot \binom{4}{1}\log\rho_1 +2 \cdot \binom{4}{2}\log\rho_0\right]\,.
	\end{equation}
	
This, in general, is different from zero. Indeed, a model with two- and four-body interactions shares the same partition $\mathcal{Q}^*$ in Equation (\ref{eq:4spin_part}). Therefore, unlike the example of two spins, symmetries of the $\mathcal{Q}^*$ model do not allow one to recover uniquely the generative model (Figure \ref{fig:Spurious}, left). Rather,~the~inferred model has a fourth order interaction (Figure \ref{fig:Spurious}, right) that cannot be excluded on the basis of symmetries alone. Note that there are $2^{2^4-1}$ = 32,768 possible models of four spins. In this case, symmetries allow us to reduce the set of possible models to just two.

	\subsection{Exchangeable Spin Models}
	\label{ex1}
	
	Consider models where $P(\spin)$ is invariant under any permutation $\pi$ of the spins, i.e., $P(s_1,\ldots,s_n)=P(s_{\pi_1},\ldots,s_{\pi_n})$. For these models, $P(\spin)$ 
only depends on the total magnetization $\sum_i s_i$. For example, the fully-connected Ising model:
\begin{equation}
\label{mfIsing}
P(\spin)=\frac{1}{\partf}e^{\frac{\beta}{n} \sum_{i<j} s_is_j}
\end{equation}
belongs to this class. 
It is natural to consider the partition where $Q_q$ contain all configurations with $q$ spins $s_i=-1$ and $n-q$ spins $s_j=+1$ ($q=0,1,\ldots,n$). Therefore, when computing $\chi_q^\mu$, one has to consider $|Q_q|={n\choose q}$ configurations. If $\mu$ involves $m$ spins, then ${m\choose j}{n-m\choose q-j}$ of them will involve $j$ spins $s_i=-1$, and the operator $\phi^\mu(s)$ takes the value $(-1)^j$ on these configurations. Therefore, $\chi_q^\mu$ only depends on the number $m=|\mu|$ of spins involved and:
\[
\chi_q^\mu=\chi_{q,m}\equiv \frac{1}{2^n}\sum_{j=0}^q {m\choose j}{n-m\choose q-j}(-1)^j,\qquad m=|\mu|.
\]

This implies that the coefficients $g^\mu$ of terms that involve $m$ spins must all be equal. Indeed,~for~any two operators $\mu\neq\mu'$ with $|\mu|=|\mu'|$:
\begin{equation}
\label{ }
g^\mu=\sum_{q=0}^n\chi_{q,m}\log\mu_q=g^{\mu'}.
\end{equation}

Therefore, the proposed scheme is able, in this case, to reduce the dimensionality of the inference problem dramatically, to models where interactions $g^\mu$ only depend on the number $m=|\mu|$ of spins involved in $\phi^\mu$.

Note also that any non-null vector $v^\mu$ such that $\sum_\mu v^\mu=0$ and $v^\mu=0$ if $|\mu|\neq m>0$ satisfies:
\[
\sum_{\mu}v^\mu\chi^\mu_q=0.
\]

The vectors $\vecu_\lambda$ corresponding to the non-zero singular values of $\hat \chi$ need to be orthogonal to each of these vectors, so they need to be constant for all $\mu$ that involve the same number of spins. In other words, $u^\mu_\lambda=u_\lambda(|\mu|)$ only depend on the number $|\mu|$ of spins involved. A suitable choice of a set of $n$~independent eigenvectors in this space is given by $u_\lambda(m)=a\delta_{\lambda,m}$ that correspond to vectors that are constant within the sectors of $\mu$ with $|\mu|=\lambda$ and are zero outside. In such a case, the sufficient statistics for models of this type are: 
\[
\psi_\lambda(\spin)=\delta_{\sum_i s_i,n-2\lambda}\,,
\]
as it should indeed be. We note in passing that terms of this form have been used in \cite{TkaciketalJStatMech2013}. 

Inference can also be carried out directly. We first observe that the $g_\lambda$ are defined up to a constant. This allows us to fix one of them arbitrarily, so we will take $g_0=0$. If $K_\lambda$ is the number of observed configurations with $\lambda$ spins $s_i=-1$, then the equation $\langle\phi_\lambda\rangle=\overline{\phi_\lambda}$ (for $\lambda>0$) reads:
\[
{n\choose \lambda}\frac{e^{g_\lambda}}{\partf}=\frac{K_\lambda}{N},\qquad \partf=1+\sum_{\lambda=1}^n{n\choose \lambda} e^{g_\lambda}
\]
so that, after some algebra, 
\[
g_\lambda=\log\left(\frac{K_\lambda}{{n\choose \lambda}K_0}\right).
\]

From this, one can go back to the couplings of operators $g^\mu$ using:
\[
g^\mu
=\sum_{\lambda=0}^n
\log\left(\frac{K_\lambda}{{n\choose \lambda}K_0}\right)
\frac{1}{2^n}\sum_{j=0}^\lambda {|\mu|\choose j}{n-|\mu|\choose \lambda-j}(-1)^j
\]

\begin{figure}[H]
	\centering
		\includegraphics[width=14.5cm]{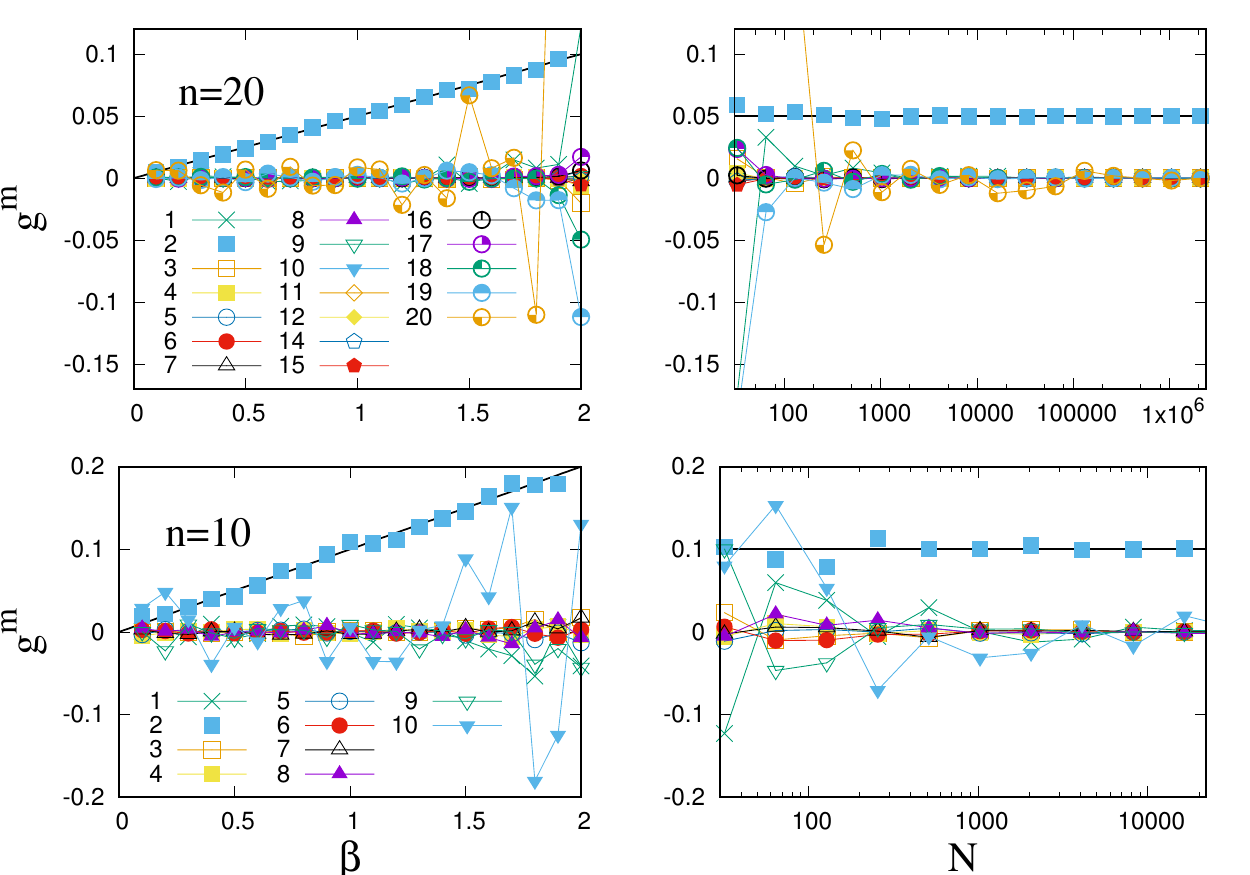}
	\caption{ \label{fig:mfIsing} Inferred parameters of a mean field Ising model Equation~(\ref{mfIsing}) with $n=10$ ({bottom}) and $20$ ({top})~spins (only interactions up to $m=10$ spins are shown for $n=20$). ({Left}) Couplings $g^m$ of \emph{m}-th order interactions as a function of $\beta$ ($N=10^3$ samples for $n=10$ and $N=10^4$ for $n=20$). ({Right}) $g^m$ as a function of $N$ for $\beta=1$. The correct value $g^2=\beta/n$ is shown as a full line.} 
\end{figure}

Figure \ref{fig:mfIsing} illustrates this procedure for the case of the mean field (pairwise) Ising model Equation~(\ref{mfIsing}). As this shows, the procedure outlined above identifies the right model when the number of samples is large enough. If $N$ is not large enough, large deviations from theoretical results start arising in couplings of highest order, especially if $\beta$ is large. 

	\subsection{The Deep Under-Sampling Limit}
	
  The case where the number $N$ of sampled configurations is so small that some of the configurations are never observed deserves some comments. As we have seen, taking the frequency partition $\mathcal{K}$, where~$Q_k=\{\spin:~k_{\spin}=k\}$, if $Q_0\not =\emptyset$, then divergences can manifest in those couplings where $\chi_0^\mu\neq 0$.
  
It is instructive to consider the deep under-sampling regime where the number $N$ of visited configurations is so small that configurations are observed at most once in the sample. This occurs when $N\ll 2^n$. In this case, the most likely partitions are {(i)} the one where all states have the same probability $\mathcal{Q}_0$ and {(ii)} the one where states observed once have probability $\rho/N$ and states not yet observed have probability $(1-\rho)/(2^n-N)$, i.e., $\mathcal{Q}_1=\{Q_0,Q_1\}$ with $Q_k=\{\spin :k_{\spin}=k\}$, $k=0,1$. Following \cite{marsilihaimovici}, it is easy to see that that generically, the probability of model $\mathcal{Q}_0$ overweights the one of model $\mathcal{Q}_1$, because $P\{\hat s|\mathcal{Q}_1\}\ll P\{\hat s|\mathcal{Q}_0\}$. Under $\mathcal{Q}_0$, it is easy to see that $\chi_0^\mu=0$ for all $\mu>0$. This, in turn, implies that $g^\mu=0$ exactly for all $\mu>0$. We reach the conclusion that no interaction can be inferred in this case \cite{footnote3}.
  
  Taking instead the partition $\mathcal{Q}_1$, a straightforward calculation shows that Equation (\ref{eq:mapp}) leads to $g^\mu=a\overline{\phi^\mu}$. Here, $a$ should be fixed in order to solve Equation (\ref{eq:inference}). It is not hard to see that this leads to $a\to\infty$. This is necessary in order to recover empirical averages, which are computed assuming that unobserved states $\spin\in Q_0$ have zero probability. 
  
  This example suggests that the divergences that occur when assuming the $\mathcal{K}$ partition, because~of unobserved states ($k_{\spin}=0$) can be removed by considering partitions where unobserved states are clamped together with states that are observed once. 
  
  These singularities arise because, when all the singular values are considered, the maximum entropy distribution {exactly} reproduces the empirical distribution. This suggests that a further method to remove these singularities is to consider only the first $\ell$ singular values (those with largest $\Lambda_\lambda$) and to neglect the others, i.e., to set $g_\lambda=0$ for all other $\lambda$'s. It is easy to see that this solves the problem in the case of the deep under-sampling regime considered above. There, only one singular value exists, and when this is neglected, one derives the result $\hat{g}^\mu=0$, $\forall \mu$ again. In order to illustrate this procedure in a more general setting, we turn to the specific case of the U.S. Supreme Court data \cite{BialekUSSC}.
  		
   \subsection{A Real-World Example}

We have applied the inference scheme to the data of \cite{BialekUSSC} that refer to the decisions of the U.S. Supreme Court on 895 cases. The U.S. Supreme Court is composed of nine judges, each of whom casts a vote against ($s_i=-1$) or in favor ($s_i=+1$) of a given case. Therefore, this is a $n=9$ spin system for which we have $N=895$ observations. The work in \cite{BialekUSSC} has fitted this dataset with a fully-connected pairwise spin model. We refer to \cite{BialekUSSC} for details on the dataset and on the analysis. The question we wish to address here is whether the statistical dependence between judges of the U.S. Supreme Court can really be described as a pairwise interaction, which hints at the direct influence of one judge on another one, or~whether higher order interactions are also present. 

In order to address this issue, we also studied a dataset $n=9$ spins generated form a pairwise interacting model, Equation (\ref{mfIsing}), from which we generated $N=895$ independent samples. The value of $\beta=2.28$ was chosen so as to match the average value of two-body interactions fitted in the true dataset. This allows us to test the ability of our method to recover the correct model when no assumption on the model is made. 

As discussed above, the procedure discussed in the previous section yields estimates $\hat g^\mu$ that allow us to recover empirical averages of {all} the operators. 
These, for a finite sample size $N$, are likely to be affected by considerable noise that is expected to render the estimated $\hat g^\mu$ extremely unstable. In~particular, since the sample contains unobserved states, i.e., states with $k_{\spin}=0$, we expect some of the parameters $g^\mu$ to diverge or, with a finite numerical precision, to attain large values.

Therefore, we also performed inference considering only the components with largest $\Lambda_\lambda$ in the singular value decomposition. Table \ref{tab1} reports the values of the estimated parameters $\hat g_\lambda$ obtained for the U.S. Supreme Court considering only the top $\ell=2$ to $7$ singular values, and it compares them to those obtained when all singular values are considered. We observe that when enough singular values are considered, the estimated couplings converge to stable values, which are very different from those obtained when all 18 singular values are considered. This signals that the instability due to unobserved states can be cured by neglecting small singular values $\Lambda_\lambda\ll 1$.

\begin{table}[htp]
\begin{center}
\begin{tabular}{cccccccc}
\toprule
\boldmath{$\lambda$} & \boldmath{$\Lambda_\lambda$} & \boldmath{$\hat g_\lambda^{(2)}$} & \boldmath{$\hat g_\lambda^{(3)}$} & \boldmath{$\hat g_\lambda^{(4)}$} & \boldmath{$\hat g_\lambda^{(5)}$} & \boldmath{$\hat g_\lambda^{(7)}$} & \boldmath{$\hat g_\lambda$} \\
\midrule
1 & 0.528 & 0.946 	& 1.023 	& 1.347 	& 1.512 	& 1.510 	& 3.680 \\
2 & 0.250 & $-$0.506 	& $-$0.573 	& $-$0.688 	& $-$0.722 	& $-$0.722 	& $-$1.213 \\
3 & 0.159 & 0 		& 0.256 	& 0.358	& 0.378 	& 0.377 	& 0.519 \\
4 & 0.102 & 0 		& 0 		& $-$0.436 	& $-$0.492 	& $-$0.491 	& $-$0.601\\
5 & 0.073 & 0 		& 0 		& 0 	& $-$0.178 	& $-$0.131 	& $-$0.152 \\
6 & 0.062 & 0 		& 0 		& 0 	& 0 	& 0.018 	& 0.087\\
7 & 0.062 & 0 		& 0 		& 0 	& 0 	& $-$0.010 	& $-$0.041 \\
8 & 0.055 & 0 		& 0 		& 0 	& 0 	& 0 	& $-$0.222 \\

\bottomrule
\end{tabular}
\caption{\label{tab1} Singular values and estimated parameters for the U.S. Supreme Court data. The parameters $\hat g^{(\ell)}_\lambda$ refer to maximum entropy estimates of the model that considers only the top $\ell$ singular values (i.e.,~$\lambda\le \ell$), whereas $\hat g_\lambda$ in the last column refers to estimated parameters using all singular values. }
\end{center}
\end{table}%

\begin{figure}[H]
	\centering
		\includegraphics[width=12cm]{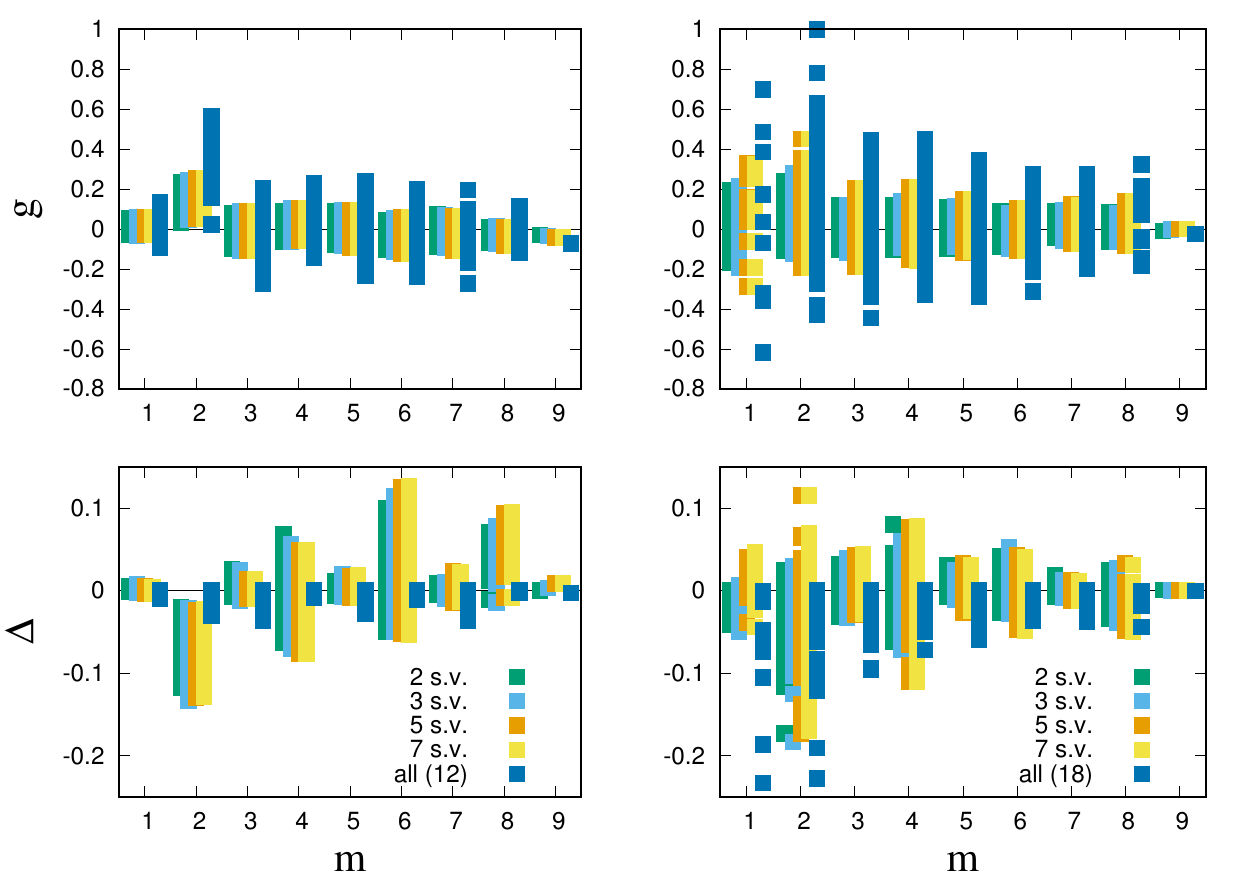}
	\caption{ \label{fig:ussc} Inference of a system of $n=9$ spins from a dataset of $N=895$ samples. ({Left}) Data generated from a pairwise Ising model Equation (\ref{mfIsing}) with $\beta=2.28$. ({Right}) Data from the U.S. Supreme Court \cite{BialekUSSC}. The upper panels report the estimated values of the parameters $\hat g^\mu$ as a function of the order $m$ of the interaction. Different colors refer to inference limited to the largest $\ell=2,3,5,7$ singular values or to the case when all singular values are considered. The lower panels report the change in log likelihood (per sample point) when a single parameter $g^\mu$ is set to zero, as a function of the order $m=|\mu|$ of the interaction.}
\end{figure}

This is confirmed by Figure \ref{fig:ussc}, which shows that estimates of $\hat g^\mu$ are much more stable when few singular values are considered (top right panel). The top left panel, which refer to synthetic data generated from Equation (\ref{mfIsing}), confirms this conclusion. The estimates $\hat g^\mu$ are significantly larger for a two-body interaction than for higher order and one-body interactions, as expected. Yet, when all singular values are considered, the estimated values of a two-body interaction fluctuate around values that are much larger than the theoretical one ($\beta/n\simeq 0.2533\ldots$) and the ones estimated from fewer singular values.

In order to test the performance of the inferred couplings, we measure for each operator $\mu$~the~change: 
\begin{equation}
\label{ }
\Delta_\mu=\max_{\vecg: g^\mu=0}\sum_\mu g^\mu\overline{\phi^\mu}-\sum_\mu \hat g^\mu\overline{\phi^\mu}
\end{equation}
in log-likelihood when $g^\mu$ is set to zero. If $\Delta_\mu$ is positive or is small and negative, the coupling $g^\mu$ can be set to zero without affecting much the ability of the model to describe the data. A large and negative $\Delta_\mu$ instead signals a relevant interaction $g^\mu$.

Clearly, $\Delta_\mu\le 0$ for all $\mu$ when $\hat g^\mu$ is computed using all the $q$ components. This is because in that case, the log-likelihood reaches the maximal value it can possibly achieve. When not all singular values are used, $\Delta_\mu$ can also attain positive values. 

Figure \ref{fig:ussc} confirms our conclusions that inference using all the components is unstable. Indeed~for the synthetic data, the loss in likelihood is spread out on operators of all orders, when all singular values are considered. When few singular values are considered, instead, the loss in likelihood is heavily concentrated on two body terms (Figure \ref{fig:ussc}, bottom left). Pairwise interactions stick out prominently because $\Delta_\mu<0$ for all two-body operators $\mu$. Still, we see that some of the higher order interactions, with even order, also generate significant likelihood losses. 

\begin{figure}[H]
\centering
		\includegraphics[width=0.5\textwidth]{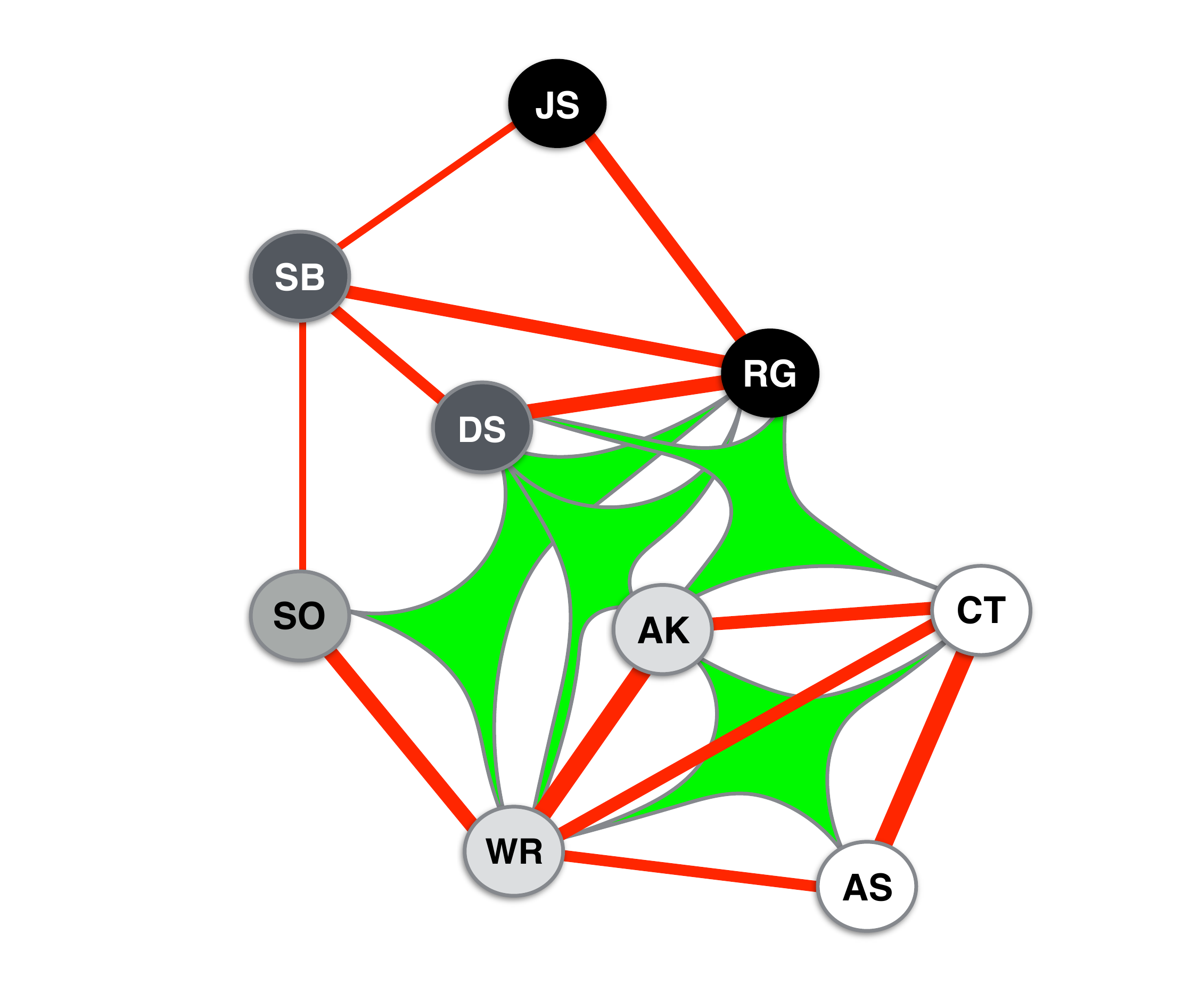}
	\vspace{-6pt}

	\caption{ \label{fig:ussc_net} Hypergraph of the top 15 interactions between the nine judges of the second Rehnquist Court. Judges are represented as nodes with labels referring to the initials (as in \cite{BialekUSSC}). Two-body interactions are represented by (red) links of a width that increases with $|\Delta_\mu|$, whereas four-body interactions as (green) shapes joining the four nodes. The shade of the nodes represents the ideological orientation, as~reported in \cite{BialekUSSC}, from liberal (black) to conservative (white).}
\end{figure}

With this insight, we can now turn to the U.S. Supreme Court data, focusing on inference with few singular values. Pairwise interactions stick out as having both sizable $\hat g^\mu$ (Figure \ref{fig:ussc}, top right) and significant likelihood loss (Figure \ref{fig:ussc}, bottom right). Indeed, the top interactions (those with minimal $\Delta_\mu$) are prevalently pairwise ones. Figure \ref{fig:ussc_net} shows the hypergraph obtained by considering the top 15~interactions \cite{footnote4}, which are two- or four-body terms (see the caption for details). Comparing this with synthetic data, where we find that the top 19 interactions are all pairwise, we conjecture that four-body interactions may not be spurious. 
The resulting network clearly reflects the orientation of individual judges across an ideological spectrum going from liberal to conservative positions (as~defined in \cite{BialekUSSC}). Interestingly, while two-body interactions describe a network polarized across this spectrum with two clear groups, four-body terms appear to mediate the interactions between the two groups. 
The prevalence of two-body interactions suggests that direct interaction between the judges is clearly important, yet higher order interactions seem to play a relevant role in shaping their collective behavior.

As in the analysis in \cite{BialekUSSC}, single-body terms, representing {a priori} biases of individual judges, are not very relevant \cite{footnote5}.	
   	
	\section{Conclusions}
		
   The present work represents a first step towards a Bayesian model selection procedure for spin models with interactions of arbitrary order. Rather than tackling the problem directly, which would imply comparing an astronomical number of models even for moderate $n$, we show that model selection can be performed first on mixture models, and then, the result can be projected in the space of spin models. This approach spots {symmetries} between states that occur with a similar frequency, which~impose constraints between the parameters $g^\mu$. As we have seen, in simple cases, these symmetries are enough to recover the correct sparse model, imposing that $g^\mu=0$ for all those interactions $\phi^\mu$ that are not consistent with the symmetries. 
   These symmetries allow us to derive a set of sufficient statistics $\psi_\lambda(\spin)$ (the relevant variables) whose empirical values allow one to derive the maximum likelihood parameters $\hat g_\lambda$. The number $q$ of sufficient statistics is given by the number of sets in the optimal partition $\mathcal{Q}^*$ of states. Therefore, the dimensionality of the inference problem is not related to the number of different interaction terms $\phi^\mu(\spin)$ (or equivalently, of parameters $g^\mu$), but it is rather controlled by the number of different frequencies that are observed in the data. As the number $N$~of samples increases, $q$ increases and so does the dimensionality of the inference problem, until one reaches the well-sampled regime ($N\gg 2^n$) when all states $\spin$ are well {resolved} in frequency.
   
   	It has been observed \cite{Amari3} that the family of probability distributions of the form (\ref{eq:1}) is endowed with a hierarchical structure that implies that high-order and low-order interactions are entangled in a nontrivial way. 
	For example, we observe a non-trivial dependence between two- and four-body interactions. On the other hand, \cite{Beretta} shows that the structure of interdependence between operators in a model is not simply related to the order of the interactions and is invariant with respect to {gauge transformations} that do not conserve the order of operators. This, combined with the fact that our approach does not introduce any explicit bias to favor an interaction of any particular order, suggests~that the approach generates a genuine prediction on the relevance of interactions of a particular order (e.g., pairwise). 
	Yet, it would be interesting to explore these issues further, combining~the quasi-orthogonal decomposition introduced in \cite{Amari3} with our approach.
   
It is interesting to contrast our approach with the growing literature on sloppy models (see, e.g., \cite{sloppy}). Transtrum et al. \cite{sloppy} have observed that inference of a given model is often plagued by overfitting that causes large errors in particular combinations of the estimated parameters. 
	 
		Our approach is markedly different in that we stem right from the beginning from Bayesian model selection, and hence, we rule out overfitting from the outset. Our decomposition in singular values identifies those directions in the space of parameters that allow one to match the empirical distribution while preserving the symmetries between configurations observed with a similar frequency. 
		
The approach discussed in this paper is only feasible when the number of variables $n$ is small. Yet,~the~generalization to a case where the set $\mathcal{M}$ of interactions is only a subset of the possible interactions is straightforward. This entails setting to zero all couplings $g^\mu$ relative to interactions $\mu\not\in\mathcal{M}$. Devising decimation schemes for doing this in a systematic manner, as well as combining our approach with regularization schemes (e.g., LASSO) to recover sparse models comprise a promising avenue of research for exploring the space of models.

\vspace{6pt} 


\acknowledgments{ We gratefully acknowledge Edward D. Lee and William Bialek for sharing the data of \cite{BialekUSSC}. We are grateful to Iacopo Mastromatteo, Vijay Balasubramanian, Yasser Roudi, Cl\'elia de Mulatier and Paolo Pietro Mazza for insightful discussions.}

\authorcontributions{L.G. and M.M. conceived the research, performed the analysis and wrote the paper.}

\conflictsofinterest{The authors declare no conflict of interest.}
\appendixtitles{yes}
\appendix
\appendixsections{multiple} 
\section{The Completeness Relation (\ref{eq:mapp})}
\label{appendix:comp}

We notice that the set of operators satisfies the following orthogonality relations:
	\begin{align}
	&\label{eq:ort1} \sum_{\{\spin\}} \phi^\mu(\mathbf{s})\phi^\nu(\mathbf{s}) = \delta_{\mu, \nu}\cdot 2^n; \\
	&\label{eq:ort2} \sum_{\{\spin\}} \phi^\mu(\mathbf{s}) = 0\,,\qquad \forall \mu>0\,.
	\end{align}
	
	Then, taking the logarithm of Equation (\ref{eq:1}), multiplying by $\phi^\nu(\spin)/2^n$ with $\nu>0$ and summing over $\spin$, one finds that:
	\begin{align}
	& \frac{1}{2^n} \sum_{\{\spin\}} \phi^\nu(\mathbf{s}) \cdot \left( \sum_\mu \phi^\mu(\mathbf{s}) g^\mu - \log \mathcal{Z} \right)\\
	&= \sum_\mu \delta_{\mu, \nu} g^\mu = g^\nu \,,
	\end{align}
	
	Combining the above identity with the expression with Equation (\ref{mixturecomb}) finally yields Equation (\ref{eq:mapp}). 
%
%
%
%
	
  \section{The Posterior Distribution of $\varrho$}
  \label{Dirichlet}
  
  Following \cite{marsilihaimovici}, we assume a Dirichlet prior for the parameter vector $\vecrho$, i.e.,
\begin{equation}
P_0^{\mathcal{Q}}(\vecrho)=\frac{\Gamma(a\cdot2^n)}{\left[\Gamma(a)\right]^{2^n}}\prod_i \rho_i^{a-1} \delta\left(\sum_i \rho_i -1\right)\,.
\end{equation}

This is a natural choice due to the fact that it is a conjugate prior for the parameters of the multinomial distribution \cite{gelman}. This means that the posterior distribution has the same functional form as the prior, and the $a$ parameters can be interpreted as pseudocounts. 
In other words, the posterior probability of $\vec{\rho}$ is still a Dirichlet distribution, i.e.,
\begin{equation}
\label{posterior}
P_1^{\mathcal{Q}}(\vecrho|\hat{s})=\Gamma\left(\sum_i k_i +a \right)\prod_i\frac{ \rho_i^{k_i+a-1}}{\Gamma(k_i+a)} \delta\left(\sum_i \rho_i -1\right)\,,
\end{equation}
where $k_s$ is the number of times state $s_i$ was observed in the sample. We remind that the choice $a=0.5$ corresponds to the least informative (Jeffrey's) prior \cite{box}, whereas with $a=0$, the expected values of the parameters $\varrho_q$ coincide with the maximal likelihood estimates. 

  The likelihood of the sample $\hat s$ under model $\mathcal{Q}$ is given by:
  \[
  P\{\hat s|\mathcal{Q}\}=\sum_j\left[\log\frac{\Gamma (K_j+a)}{\Gamma(a)}-K_j\log m_j\right]-\log\frac{\Gamma(N+aq)}{\Gamma(aq)}
  \]
  where: 
  \[
  K_j=\sum_{\spin\in Q_j}k_{\spin},
  \]
  is the number of sample points in partition $Q_j$ and $m_j=|Q_j|$ is the number of states in partition $Q_j$. The work in \cite{marsilihaimovici} shows that the partition that maximizes the likelihood is the one where states with similar frequencies are grouped together, which is a coarse-graining of the $\mathcal{K}$ partition.

	\section{The Covariance Matrix of the $g^\mu$ Parameters}
  \label{covarmat}
	The covariance matrix $\hat{C}$ can be written as:
	\begin{equation}
	\label{covg}
	C^{\mu \nu}=\mathrm{Cov}[g^\mu, g^\nu]=\frac{1}{2^{2n}}\sum_{q,q'} \chi_q^\mu \chi_{q'}^\mu \mathrm{Cov}[\log\rho_q, \log\rho_{q'}]\,.
	\end{equation}
	
	In order to compute the covariance of the $\vecg$ parameters, $\mathrm{Cov}[\log \rho_q, \log \rho_{q'}]$ must be computed first. For the following considerations, it is useful to define:
\begin{equation}
\label{eq:A8}
\mathcal{Z}(\vec{\lambda})=\mathbf{E}\left[\prod_q \rho_q^{\lambda_q} \right]\,,
\end{equation}
where the expectation is taken over the posterior distribution of $\vec{\rho}$:
\begin{equation}
\label{posteriordir}
P_1^{\mathcal{Q}}(\vec{\rho}|\hat{s})=\Gamma\left(\sum_i k_i +a \right)\prod_i\frac{ \rho_i^{k_i+a-1}}{\Gamma(k_i+a)} \delta\left(\sum_i \rho_i -1\right)\,,
\end{equation}
as in \cite{marsilihaimovici}.
	First, one can see that:
	\begin{align*}
	\partial_{\lambda_q} \partial_{\lambda_{q'}} \mathcal{Z}(\vec{\lambda}) &=\mathbf{E}\left[ \delta_{q,q'}\cdot \rho_q^{\lambda_q} (\log \rho_q)^2 \prod_{q \neq q''} \rho_{q''}^{\lambda_{q''}}\right] +\\
	&+ \mathbf{E}\left[(1-\delta_{q,q'}) \cdot \rho_q^{\lambda_q} \log \rho_q \cdot \rho_{q'}^{\lambda_{q'}} \log \rho_{q'} \prod_{q \neq q'', q'} \rho_{q''}^{\lambda_{q''}}\right]\,.
	\end{align*}
This relation, for $\vec{\lambda} = 0$, yields:
	\begin{align}
	\partial_{\lambda_q} \partial_{\lambda_{q'}} \mathcal{Z}(\vec{\lambda})\Bigr|_{\vec{\lambda}=0}&= \mathbf{E}\left[ \delta_{q,q'}\cdot (\log \rho_q)^2 + (1-\delta_{q,q'}) \cdot \log \rho_q \cdot \log \rho_{q'} \right] \\
	&= \mathbf{E}\left[ \log \rho_q, \log \rho_{q'}\right]\,,
	\end{align}
	and we find that:
	\begin{align}
	\mathrm{Cov}[\log \rho_q, \log \rho_{q'}]&=\left[\partial_{\lambda_q} \partial_{\lambda_{q'}} \mathcal{Z}(\vec{\lambda}) -\partial_{\lambda_q} \mathcal{Z}(\vec{\lambda}) \partial_{\lambda_{q'}}\mathcal{Z}(\vec{\lambda})\right]\Bigr|_{\vec{\lambda}=0}\\
	&=\frac{\partial^2}{\partial_\lambda \partial_{\lambda'}} \left[ \log \mathcal{Z}(\vec{\lambda}) \right]\Bigr|_{\vec{\lambda}=0}
	\end{align}
		Given the distribution of $\vec{\rho}$, the expression in Eq. (\ref{eq:A8}) reads:
	\begin{equation}\mathcal{Z}(\vec{\lambda})= \frac{1}{\prod_q m_q^\lambda}\cdot\frac{\prod_q \Gamma(K_q+\lambda_q+a)}{ \Gamma(\sum_q (K_q+\lambda_q+a))}\,.\end{equation}
then:
	\begin{align*}
	\frac{\partial^2}{\partial_\lambda \partial_{\lambda'}} \left[ \log \mathcal{Z}(\vec{\lambda}) \right]&=\frac{\partial^2}{\partial_\lambda \partial_{\lambda'}} \left[ \sum_q \log \Gamma(K_q+\lambda_q+a) - \log \Gamma(\sum_q K_q + \sum_q \lambda_q + \sum_q a) \right] \\
	&= \frac{\partial}{\partial_{\lambda'}} \left[ \frac{\Gamma'(K_q+\lambda_q+a)}{\Gamma(K_q+\lambda_q+a)} - \frac{\Gamma'(M + \sum_q \lambda_q + Na)}{\Gamma(M + \sum_q \lambda_q + N a)} \right]\\
	&=\delta_{q,q'} \cdot \psi^{(1)}(K_q+\lambda_q+a) - \psi^{(1)}(M+\sum_q \lambda_q +Na)\,
	\end{align*}
where $\psi^{(n)}(x)=\frac{d^{n+1}}{dx^{n+1}} \log \Gamma(x)$ is the polygamma function.
	
	Evaluated in $\vec{\lambda} = 0$, this yields:
	\begin{equation}
	\label{covmatr}
	\mathrm{Cov}[\log \rho_q, \log \rho_{q'}]=C_{q,q'}=\delta_{q,q'} \cdot \psi^{(1)}(K_q+a) - \psi^{(1)}(M +Na)\,.
	\end{equation}
	
	
	Therefore, the covariance matrix among the elements of $\vec{\rho}$ is composed of a diagonal part plus a part proportional to the identity matrix.
	
	Inserting (\ref{covmatr}) into (\ref{covg}), 
	\begin{align}
	\label{element}
	C^{\mu \nu}&=\mathrm{Cov}[g^\mu, g^\nu] = \frac{1}{2^{2n}}\sum_{q} \chi_q^\mu \chi_{q}^\nu \psi^{(1)}(K_q+a)-\frac{1}{2^{2n}}\cdot \psi^{(1)}(M +Na)\cdot\left(\sum_{q} \chi_q^\mu\right)\left(\sum_{q'} \chi_{q'}^\nu\right) \\
	&=\frac{1}{2^{2n}}\sum_{q} \chi_q^\mu \chi_{q}^\nu \psi^{(1)}(K_q+a)\,,
	\end{align}
due to the fact that, by Equation (\ref{eq:ort2}), $\sum_{q'} \chi_{q'}^\nu=0$ for all $\nu>0$.

\section{Sufficient Statistics}
\label{eigsuff}

Property (\ref{eq:gv}) can be used to give a proof of Equation (\ref{eq:maxentlamb}).

Let $v^\mu_j$, $j=1,\ldots, 2^n-1-q$ be the vectors such that $\sum_\mu v^\mu_j g^\mu=0$. With the $u_\lambda^\mu$, these constitute an orthonormal basis for the $2^n-1$ dimensional space with coordinates $g^\mu$. This implies the identity: 
\[
\delta_{\mu,\nu}=\sum_{\lambda}u^\mu_\lambda u^\nu_\lambda+\sum_jv_j^\mu v_j^\nu
\]
that can be inserted in the expression:
\begin{eqnarray}
\sum_\mu g^\mu\phi^\mu(\spin) & = & \sum_{\mu,\nu}g^\mu\phi^\nu(\spin)\delta_{\mu,\nu} \\
 & = & \sum_\lambda\left(\sum_\mu u^\mu_\lambda g^\mu\right)\left(\sum_\nu u^\nu_\lambda \phi^\nu(\spin)\right)+
 \sum_j\left(\sum_\mu v^\mu_j g^\mu\right)\left(\sum_\nu v^\nu_j \phi^\nu(\spin)\right)
\end{eqnarray}
that yields the desired result, because $\sum_\mu v^\mu_j g^\mu=0$ for all $j$.


\begin{thebibliography}{999}
\bibitem{footnote3}
\newblock Notice that other inference methods may infer non-zero interactions in this case \cite{review}. Note also that the statistics of the frequencies can be very different if one takes a subset of $n'<n$ spin, so the present approach may predict $g^\mu\neq 0$ when the same dataset is restricted to a subset of spins.
\end{thebibliography}
\reftitle{References and Notes}


\end{document}